\documentclass{cernrep} 
\usepackage{texnames}
\usepackage[T1]{fontenc}
\usepackage[bookmarks, colorlinks=true, linktoc=page, pdftex, linkcolor=black, citecolor=black, urlcolor=blue]{hyperref}
\usepackage{hepnames}
\usepackage{esint}
\usepackage{physics}
\usepackage{graphicx}
\usepackage{subcaption}
\usepackage{relsize}
\usepackage{amsmath}
\usepackage{color}
\usepackage{enumerate} 
\usepackage{tikz}
\usetikzlibrary{arrows,decorations}
\usepackage{booktabs}
\usepackage{floatflt}
\usepackage{amssymb}
\usepackage{dcolumn}
\usepackage{comment}
\usepackage{pgfplots}
\usepackage{color}
\pgfplotsset{compat=newest}
\usetikzlibrary{arrows,decorations,backgrounds,patterns,matrix,shapes,fit,calc,shadows,plotmarks,pgfplots.groupplots}
\usepackage{epstopdf}
\usepackage{fancyhdr}
\fancyhfoffset{4mm}
\fancypagestyle{ARTTITLE}{%
\fancyhf{} 
\lhead{\small{Proceedings of the 2018 CERN--Accelerator--School course on \it{Beam Instrumentation}, Tuusula, (Finland)}} 
\lfoot{Available online at \url{https://cas.web.cern.ch/previous-schools}}
\rfoot{\thepage\hspace*{3mm}}

}

\begin{document}

\title{Video Cameras used in Beam Instrumentation - an Overview}

\author {B. Walasek-Hoehne, K. Hoehne and R. Singh}

\institute{GSI Helmholtz Centre for Heavy Ion Research, Darmstadt, Germany}


\maketitle

\thispagestyle{ARTTITLE}
 
\begin{abstract}
Imaging systems have been an integral part of many beam monitors since the early days of accelerator diagnostics. 
The main application remains the observation of scintillating screens during commissioning, alignment and routine operation with the beam. 
Recorded images are often further analyzed to characterize the beam distribution 
for machine optimization.

This report provides an overview of imaging technologies and market trends of today. The image sensor, like TV tubes and solid state sensors (CCD, CMOS and CID), with particular focus on the aspects important for beam instrumentation will be discussed. Digital image acquisition as well as camera interfaces and radiation effects will be also presented. 

\end{abstract}
\maketitle 
\section{Introduction}
 Cameras are used to observe and store information transmitted by light. The intensity of the observed light is proportional to the impinging number of photons in it, and the photons energy is given by its wavelength ($E = hc/\lambda$). Cameras constitute of \textsl{optics} elements to generate an image of a distant object onto a \textsl{sensor}. This image on the \textsl{sensor} can be observed or stored.

More than 2000 years ago the Chinese scientist Mo-Tsu (470-391 BC) described the principle of the camera obscura. 
It is known that the Arabian mathematician and astronomer Alhazen (965-1040) used this principle to observe moon and stars.   
The development of the camera optics started in Italy in the 16th century. 
The physician and mathematician Hieronymus Cardanus (1501-1576) introduced a convex lens and Daniello Barbaro (1514-1570) the aperture.
Many early modern period artists used the camera obscura to create photo-realistic paintings. 
A famous example is Bernardo Bellotto, called Canaletto (1722-1780), whose detailed paintings of Warsaw were used to rebuild city old town after its complete destruction during the Second World War.
On August 19, 1839 the French Academy of Science published the daguerreotype process.
This date in known as the birth of photography. 
Photography, and later films, enables to store information transmitted by light in an easy and fast way.
With the invention of semiconductor based light sensitive sensors 50 years ago and the digital revolution in the recent decades, cameras became the ultimate tool for observing, storing, and analyzing information transmitted by light.  

Charged particle interaction with matter often results in emission of light either due to classical electromagnetic processes ~\cite{Artru} or electron transitions in atomic processes~\cite{Walasek}. From acceleration of ultra-relativistic particles, direct synchrotron radiation can be observed. Many beam diagnostic devices rely on these processes and are designed to acquire and process the information transmitted as light. The analysis of beam instrumentation data enables monitoring, optimization and further development of the accelerator and its components.

\begin{figure}[h!]
\centering
\includegraphics[width=1\textwidth]{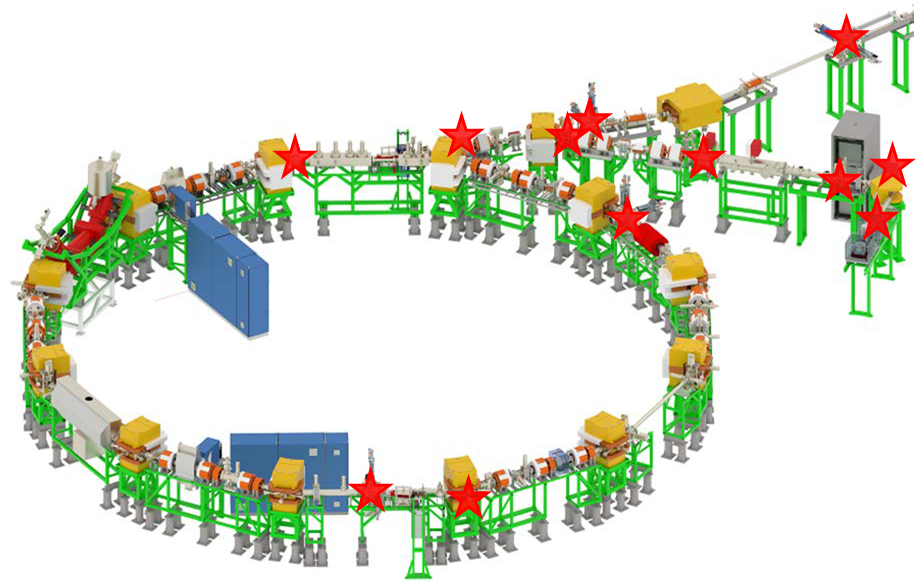}
\caption{Schematic overview of the new GSI storage ring CRYRING@ESR. Stars indicate the position of optical diagnostics station with digital camera.}
\label{fig:CRYRING}
\end{figure}

The usage of camera systems starts from monitoring the ion source (plasma region), provides the two-dimensional beam distribution measurements along the full acceleration chain as well as the first turn ring diagnostics (setup of injection and extraction) during the commissioning of a synchrotron and routine operation of the full particle accelerator facility.
We used the new GSI storage ring CRYRING@ESR with local injector shown in Figure~\ref{fig:CRYRING} to demonstrate the different task of camera system.  
At GSI approximately 80 optical diagnostic stations have been installed at various position along the whole accelerator (including ion sources, linear accelerator, ring accelerator, experimental storage rings and in High Energy Transport Lines). At the CERN facility operates more than 200 camera stations and SwissFEL is equipped with 27 transverse profile imagers for the electron beam are installed. 
 
Choosing the correct camera for a given implementation has never been a simple task, especially since applications in particle accelerators meet challenges like low/high light intensity or short/long pulses. It is not surprising that with these contradictory requirements, no single ``universal camera`` type can be applied in all the cases. The 
main challenge in imaging for beam instrumentation is still the degrading effect of the ionizing radiation on the hardware. Several investigations and developments towards radiation hard systems are still ongoing. 
This report focuses on the image sensors, radiation hardness, and acquisition and post processing. Detailed contributions on other aspects such as the source of light in beam instrumentation can be found in E. Bravin's report on Tranverse Profile measurements~\cite{Bravin3}, on the optics needed to generate the picture on the sensor in  S. Gibson's Introduction to optics ~\cite{Gibson}, and on digitization in M. Gasior's report on Analog Digital Conversion~\cite{Gasior} in the present proceedings.

\subsection{Image sensor: Video tube}
In 1908, Campbell-Swinton discussed how a fully electronic television system could be realized by using cathode ray tubes as both imaging and display devices
~\cite{Campbell}. Several different types of tubes were in use from the early 1930s to the 1980s. In these tubes, the electron beam was scanned across an image of the scene to be broadcast. The resultant current was dependent on the brightness of the image on the target. 

\begin{figure}[h!]
\centering
\includegraphics[width=1\textwidth]{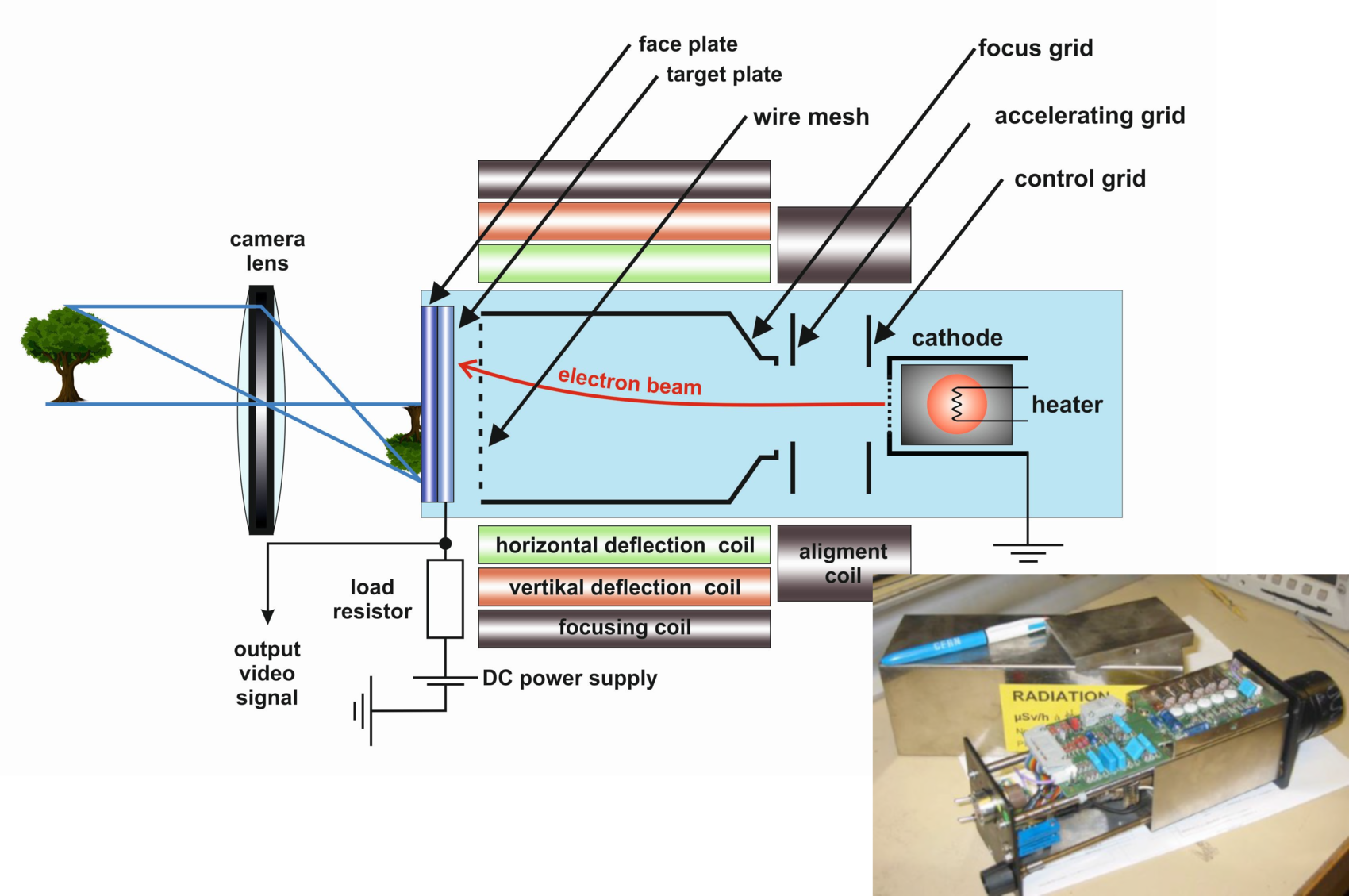} 
\caption{Vidicon camera tube schematic and hardware used at CERN SPS ring \cite{Bravin2}.}
\label{fig:Vid1}
\end{figure}
The most popular video camera tube called Vidicon was developed in the 1950s by P.K. Weimer and his colleagues in U.S. electronic company RCA~\cite{Weimer} as shown in Fig~\ref{fig:Vid1}. 
The input light from a scene passes through a lens system and hits the face plate made of glass. 
The light from the face plate falls on the so called target plate which has two layers. 
The first layer is electrically conductive, for example, a thin coat of indium oxide which is transparent to light and is a good conductor of electricity. 
It is called the signal layer. The second photosensitive layer consists of semiconducting materials such as selenium, arsenic or tellurium and is called target layer.  In the absence of light, the material is almost an insulator, having very few free electrons
which present a resistance of 20\,M$\Omega$. When light falls on it, free electrons are produced and reduces the resistance down to 2\,M$\Omega$. 

The signal layer is kept positive by an external power supply. 
The free electrons produced by incident light migrate to the positive signal layer, thus leaving a deficiency of electrons on the target layer towards gun side. 
The charge image proportional to the optical image is formed on the semiconductor layer.
When the focused electron beam produced by the electron gun is scanned through the target material, it will leave electrons in the material just sufficient to neutralize the deficiency of electrons, which results in an electric current through the load. 
This current at any spot of scanning is proportional to the number of electrons deposited by the beam which in turn is proportional to the brightness. 
The output across the load resistor is the video signal. The scanning mechanism codes the two-dimensional spatial information on brightness in a one-dimensional time dependent voltage signal.

The switch-over to solid state imaging sensors made video camera tubes obsolete for most application. However even today, Vidicon tubes are used for special tasks such as imaging in high radiant environments like nuclear power plants and accelerator facilities (e.g. SPS at CERN, see Figure~\ref{fig:Vid1}). 
The ability of these cameras to sustain very high radiation levels (tens of MRad)
is due to their composition of only passive electronics elements and valves~\cite{Bravin}, and their lifetime is estimated to be between 5.000 and 20.000 hours. The tube cameras suffer from external magnetic fields, which occur in the pulsed accelerator machines and necessitate magnetic shielding~\cite{Jung}.

\section{Image sensor CCDs}
\label{sec:CCD}
Willard Boyle and George E. Smith from Bell Laboratories, USA, invented in 1969 the Charge Coupled Device (CCD). 
They were actually working on a semiconductor version of the Bubble Memory which uses small magnetized areas, so called domains or bubbles, to store information. 
By applying an external magnetic field these bubbles can be shifted to and read out at the edge of the memory.
Basically the CCD technology uses charge trapped in potential wells created at the surface of a semiconductor to store information and shifting it by moving the potential minima. 
Already in the first publication, possible applications as imaging device, as well as display device were mentioned~\cite{Boyle}. The first patent on the application of CCDs to imaging was assigned to Michael Tompsett~\cite{Tompsett,Tompsett2} in 1971. In 2009, Boyle and Smith were awarded the Nobel Prize for Physics for their invention of the CCD concept. 

\begin{figure}[h!]
\centering
\includegraphics[width=0.55\textwidth]{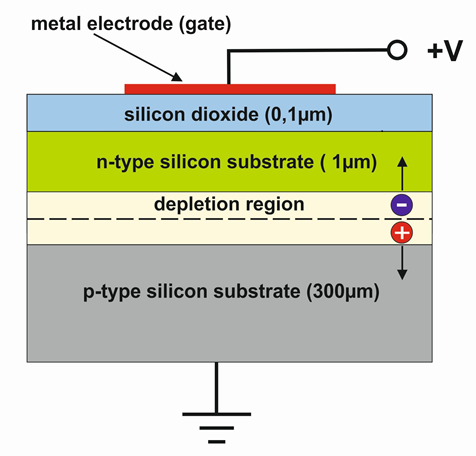} 
\caption{Scheme of the CCD sensor based on MOS capacitor.} 
\label{fig:CCD1}
\end{figure}

CCDs are manufactured using metal-oxide-semiconductor (MOS) techniques. 
Each pixel can be thought of as a MOS capacitor that converts incident photons into electrical charge and stores the charge prior to readout. 
Two different types of MOS capacitors exist: surface channel and buried channel. A typical buried channel capacitor is shown in Figure \ref{fig:CCD1} and in case of surface channel, the n-type substrate does not exist.
The two differ only slightly in their fabrication, however, nearly all manufactured CCDs use buried channel structure due to its better charge transfer efficiency. 
On top of a metal-oxide-semiconductor capacitor is a metal electrode called gate.
It is separated by an approximately 0.1\,$\mu$m thin silicon dioxide layer from an approximately 1\,$\mu$m thin n-type silicon layer which is placed on a p-type silicon substrate of approximately 300\,$\mu$m. 
A positive voltage at the gate forms the potential well directly below the electrode.
Incident light generates electron-hole pairs in the depletion zone. 
Electrons are attracted by the positive gate and migrate into the potential well.
The trapped charge in the potential well is proportional to the number of incident photons.
By changing the gate voltage of neighbouring MOS capacitors the charge can be shifted from one potential well to the next.

In general, CCD sensors use more than one MOS capacitor per pixel. 
Most common are the 3-phase sensors that have three capacitors per pixel.
CCD sensors have thousands or even millions of light sensitive pixels arranged linear (line sensors) or two-dimensional (area sensors).
Linear sensors are used e.g. in applications like flatbed scanners or fax machines.
To receive a two-dimensional image a line sensor has to be moved over the object, whereas the area sensor captures the full image within one exposure and thus is used for camera applications.
The number of pixels on a sensor reflects the development of the semiconductor industry during the past decades. 
Fairchild company produced the first commercial devices: a linear 500-pixel device and a two dimensional (100 x 100\,pixels) device. 
Nowadays, the largest commercially available camera from Seitz company has 160 million pixels (75.000 x 21.250\,pixels).

\subsection{Readout}
\label{sec:Readout} 
\begin{figure}[h!]
\centering
\includegraphics[width=0.9\textwidth]{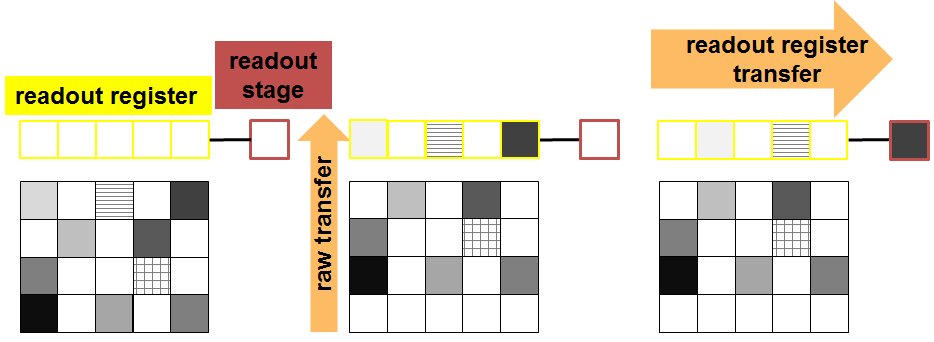} 
\caption{Readout process: during the exposure time charge proportional to the incident light is generated in each pixel (left). During the row transfer all charges of one  row of pixels are shifted towards the register (middle). Followed by a serial read out of the register (right).}
\label{fig:CCD4}
\end{figure}

Each sensor is divided in small areas, the pixels. 
During the integration time, the incoming light is converted into a proportional charge for each pixel (Figure~\ref{fig:CCD4}, left). The readout process takes place in two stages: 
the row transfer, where the charges of a row of pixels are moved in parallel into the register (Figure~\ref{fig:CCD4}, middle) followed by the second stage, the serial readout of the register, from which they are then sent to an analogue-to-digital converter (ADC) before being stored in memory~\cite{Taylor} (Figure~\ref{fig:CCD4}, right). 
As result a sequence of analogue voltage signals is generated pixel by pixel. 

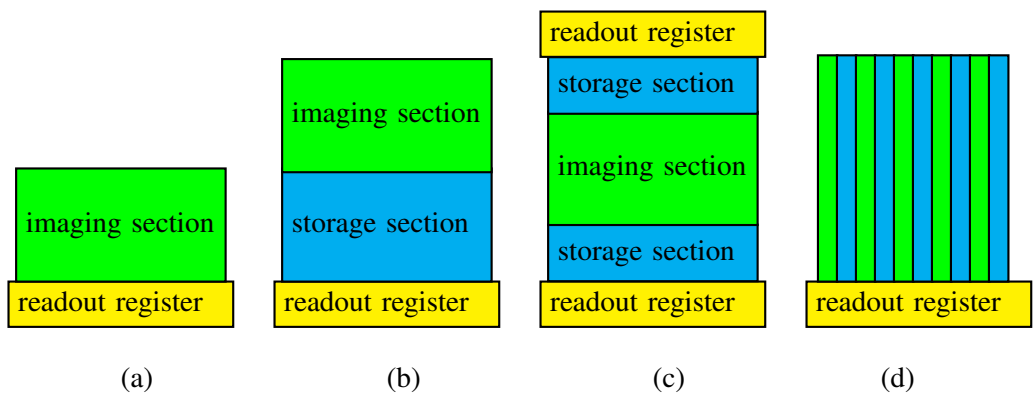
\begin{figure}
\centering
\begin{tikzpicture}
\node[thick,draw,text width=2.7cm,fill=yellow] (readout1) at (0,1) {readout register};
\node[thick,draw,text width=2.5cm,minimum height=1.5cm,fill = green] (imaging1) at (0,2.05) {imaging section};
\node[thick,text width=2cm] (label1) at (1,0) {(a)};

\node[thick,draw,text width=2.7cm,fill=yellow] (readout2) at (3.5,1) {readout register};
\node[thick,draw,text width=2.5cm,minimum height=1.5cm,fill=cyan] (storage2) at (3.5,2.05) {storage section};
\node[thick,draw,text width=2.5cm,minimum height=1.5cm,fill=green] (imaging2) at (3.5,3.5) {imaging section};
\node[thick,text width=2cm] (label2) at (4.5,0) {(b)};

\node[thick,draw,text width=2.7cm,fill=yellow] (readout3) at (7,1) {readout register};
\node[thick,draw,text width=2.5cm,minimum height=0.75cm,fill=cyan] (storage3) at (7,1.68) {storage section};
\node[thick,draw,text width=2.5cm,minimum height=1.5cm,fill=green] (imaging3) at (7,2.8) {imaging section};
\node[thick,draw,text width=2.5cm,minimum height=0.75cm,fill=cyan] (storage4) at (7,3.9) {storage section};
\node[thick,draw,text width=2.7cm,fill=yellow] (readout4) at (7,4.58) {readout register};
\node[thick,text width=2cm] (label3) at (8,0) {(c)};

\node[thick,draw,text width=2.7cm,fill=yellow] (readout3) at (10.5,1) {readout register};
\node[thick,draw,minimum width=0.25cm,minimum height=3cm,fill=green] (imaging1) at (9.3,2.8) {};
\node[thick,draw,minimum width=0.25cm,minimum height=3cm,fill=cyan] (storage1) at (9.55,2.8) {};
\node[thick,draw,minimum width=0.25cm,minimum height=3cm,fill=green] (imaging2) at (9.8,2.8) {};
\node[thick,draw,minimum width=0.25cm,minimum height=3cm,fill=cyan] (storage2) at (10.05,2.8) {};
\node[thick,draw,minimum width=0.25cm,minimum height=3cm,fill=green] (imaging3) at (10.30,2.8) {};
\node[thick,draw,minimum width=0.25cm,minimum height=3cm,fill=cyan] (storage3) at (10.55,2.8) {};
\node[thick,draw,minimum width=0.25cm,minimum height=3cm,fill=green] (imaging4) at (10.80,2.8) {};
\node[thick,draw,minimum width=0.25cm,minimum height=3cm,fill=cyan] (storage4) at (11.05,2.8) {};
\node[thick,draw,minimum width=0.25cm,minimum height=3cm,fill=green] (imaging5) at (11.30,2.8) {};
\node[thick,draw,minimum width=0.25cm,minimum height=3cm,fill=cyan] (storage5) at (11.55,2.8) {};
\node[thick,text width=2cm] (label3) at (11,0) {(d)};
\end{tikzpicture}
\caption{Area Array CCD architectures: (a) Full frame area array CCD, (b) Frame transfer area array CCD, (c) Split frame area array CCD and (d) Interline transfer area array CCD.}
\label{fig:CCD5}
\end{figure}

This sequential readout process is time consuming. 
During the readout the pixels not shifted to the register still convert incident light to charge, which can result in smearing and blurring. 
A countermeasure is using a mechanical shutter to shield the sensor from light. 
The readout time limits high-speed application as well, since the readout time shortens the available exposure time.
Several CCD architectures were developed to tackle this problems. 
The full frame sensor  (Figure~\ref{fig:CCD5}~(a)) has the architecture described above.
The frame transfer sensor  (Figure~\ref{fig:CCD5}~(b)) has a light protected storage section of the same size as the imaging section. 
Here the complete image is transferred row by row from the imaging to the storage section and then the above described readout process starts.
During the readout process the imaging section can be exposed again.
Drawback of this solution is the large reduction of the fill factor (see Section~\ref{sec:SensorCharacteristics}).
The split frame transfer architecture (Figure~\ref{fig:CCD5}~(c))  has two storage sections of half size of the imaging section and two registers. 
The time between two exposures is reduced by a factor of two compared to the frame transfer architecture.
Following this idea one ends up with the interline transfer architecture (Figure~\ref{fig:CCD5}~(d)). 
Each pixel column has a storage column next to it.
The transfer into the storage section is then one step, only.
Therefore interline transfer CCDs are fast and a mechanical shutter is not needed, but on cost of a low fill factor of typically 40\%. 

\subsection{Sensor Characteristics}
\label{sec:SensorCharacteristics} 

\begin{figure}[h!]
\centering
\includegraphics[width=0.95\textwidth]{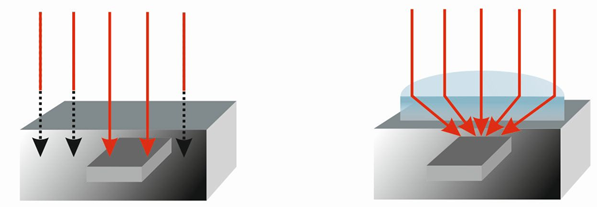}
\caption{Cross sectional view of the pixel with its photosensitive area indicated. The fill factor is approximately 40\% (left). 
The microlens on top of the pixel focus the incoming light onto the photosensitive area. The fill factor increases up to 100\% (right).}
\label{fig:CCD9}
\end{figure}

\paragraph{Fill factor}
The fill factor is the ratio of the light sensitive and the insensitive area of a sensor. 
A fill factor of 100\% is needed to capture all incident photons. 
In reality, the fill factor is smaller. 
Full frame sensors can reach the close to 100\%. 
The storage sections of frame transfer and interline transfer sensors lower the fill factor significantly.
The fill factor can be improved by placing a microlens on top of each pixel which collects and focuses the incoming light onto the sensitive area, see Figure~\ref{fig:CCD9}. The improved fill factor comes at the price of undesired optical effects,such as focussing of oblique rays into adjacent pixels. 

\paragraph{Quantum efficiency (QE)}
The quantum efficiency is the ratio of light that the sensor converts into charge. 
In case of visible light a  quantum efficiency of 60\% means, that if 10 photons hitting a pixel six electrons will be trapped in the potential well on average. 
The quantum efficiency is a sensor specific quantity. 
It depends on the design as well as on the wavelength of the incoming light. 
Some incident photons may not be absorbed due to reflection or may be absorbed where the electrons cannot be collected.
Standard CCDs are more sensitive to green and red wavelengths in the region between 550 and 900\,nm. 
Often the structures on the surface absorb the shorter wavelengths reducing the blue sensitivity of the sensor

\begin{figure}
\centering
\begin{subfigure}[b]{0.48\textwidth}
\includegraphics[width=\textwidth]{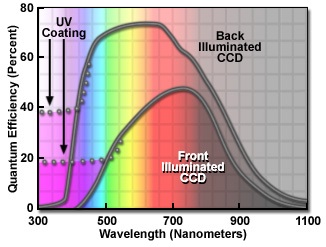}
\caption{}
\end{subfigure}
\begin{subfigure}[b]{0.5\textwidth}
\begin{tikzpicture}[scale=0.85]
\begin{semilogyaxis}[
xlabel={Temperature (K)},
ylabel={Dark current (e/pixel/s)},
xmin=200, xmax=300,
ymin=0.2, ymax=19000,
]
\addplot [semithick,mark=square,red,mark options={solid},   smooth]
coordinates {
(293.12015503875966, 10202.623822686739)
(287.984496124031, 6626.985468128054)
(283.0426356589147, 4123.316628990366)
(277.906976744186, 2678.2482482642517)
(272.9651162790698, 1702.657482042099)
(267.92635658914725, 1036.819384012191)
(262.984496124031, 591.9030770607097)
(257.8488372093023, 352.75368897697)
(253.19767441860463, 210.25087439292577)
(247.96511627906978, 117.46604264468361)
(243.02325581395348, 64.23448385710375)
(238.0813953488372, 36.67040685649364)
(233.04263565891472, 18.3983720040021)
(228.10077519379843, 9.637041027328035)
(223.25581395348837, 4.8353229656575065)
(221.1240310077519, 3.6536835347717527)
(219.08914728682169, 2.644562504820555)
(217.82945736434107, 2.225716026283415)
(215.79457364341084, 1.6818383444805345)
(212.8875968992248, 1.0468997427700588)
(208.13953488372093, 0.4924459218699044)
(202.71317829457362, 0.21250317724944057)
};
\end{semilogyaxis}
\end{tikzpicture}
\caption{}
\end{subfigure}
\caption{Wavelength dependency of the quantum efficiency for standard CCD and backside-illuminated CCD sensors \cite{Hamamatsu}. (a) Variation in dark current with temperature for a CCD~\cite{McFee}. (b) }
\label{fig:CCD20}
\end{figure}
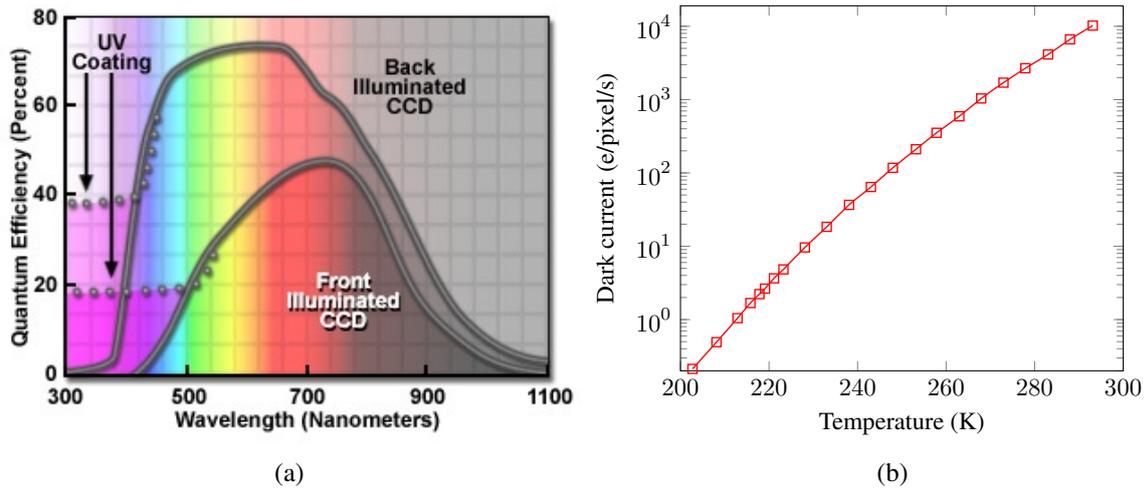

The losses due to gate structure and the insensitive layer near the surface are eliminated by the backside-illuminated CCD design, where the light falls onto the back of the CCD. 
For this type of sensors the substrate is thinned down to 10-15\,$\mu$m. 
Figure~\ref{fig:CCD20} shows the quantum efficiency as a function of the wavelength of the incoming light for standard CCD and backside-illuminated CCD senors. 
The plot shows an extended sensitivity to lower wavelength and a higher quantum efficiency over the entire range\cite{Hamamatsu}. 

\paragraph{Dark current noise}
Thermal fluctuations can generate electron-hole pairs in the pixel.  
Similar to photon generated electron-hole pairs the randomly generated electrons can be captured in the potential well, leading to noise.
This effect is called dark current noise.
The dark current's dependency on temperature follows nearly an exponential law.
So for applications requiring very low noise levels, e.g. astrophotography, cooling of the CCD chip is needed.
Since the dark current is a natural thermal process, the noise level is also influenced by the integration time and the storage time before readout. 

\begin{figure}[h!]
\centering
\begin{subfigure}[b]{0.4\textwidth}
\includegraphics[width=\textwidth]{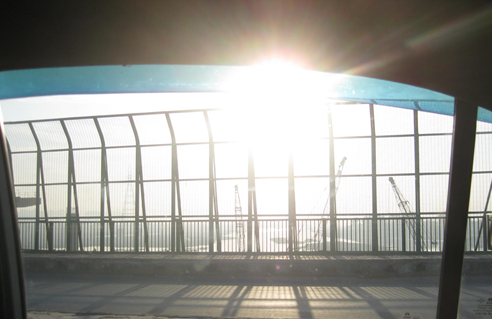}
\caption{}
\end{subfigure}
\begin{subfigure}[b]{0.55\textwidth}
\includegraphics[width=\textwidth]{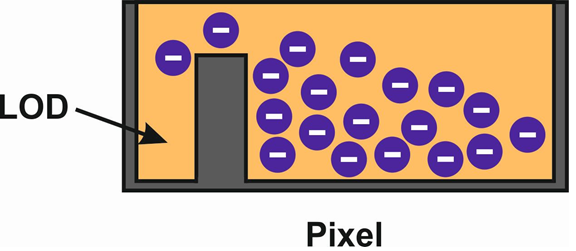}
\caption{}
\end{subfigure}
\caption{Image with blooming effect caused by excessive light level. (a) Schema of lateral overflow drain (LOD). (b)}
\label{fig:CCD12}
\end{figure}

\paragraph{Full well capacity} 
The full well capacity is the maximum number of electrons that a pixel can hold. 
Larger pixel have higher full well capacity, better signal to noise ratio (SNR), and increased dynamic range.
Typical full well capacities are 4.000 electrons for small pixels, 10.000 electrons for medium size pixels and 50.000 pixels for large pixels. 

If the number of electrons in a pixel exceeds the full well capacity, the potential well overflows. 
This leads to an increased number of electrons in the surrounding potential wells, and thus creates an area of saturated pixels. 
This effect is called blooming and is caused by bright light sources. 
Figure~\ref{fig:CCD12} shows this effect. This effect comes into CCD "by design" since the charge should be transferred to the neighbouring pixels before readout.
For quantitative beam instrumentation measurements it is important to avoid blooming, which leads to false beam profile reading.
A method to overcome blooming uses lateral overflow drains (LOD) to collect and dissipate the excess charge without affecting the surrounding pixels.
The principle of LOD is illustrated in Figure~\ref{fig:CCD12}, right.

\paragraph{Charge transfer efficiency (CTE)}
The charge transfer efficiency characterizes the CCD sensors ability to move the charge from the pixel to the output without losses. 
Nowadays CCD have typical values above 99.999\%. 

\noindent More detailed information on CCD imaging sensors is given e.g. in \cite{Taylor, PCO, PCO2}.

\section{Image sensor: CMOS}
\label{sec:CMOS}
\begin{figure}[h!]
\centering
\includegraphics[width=1\textwidth]{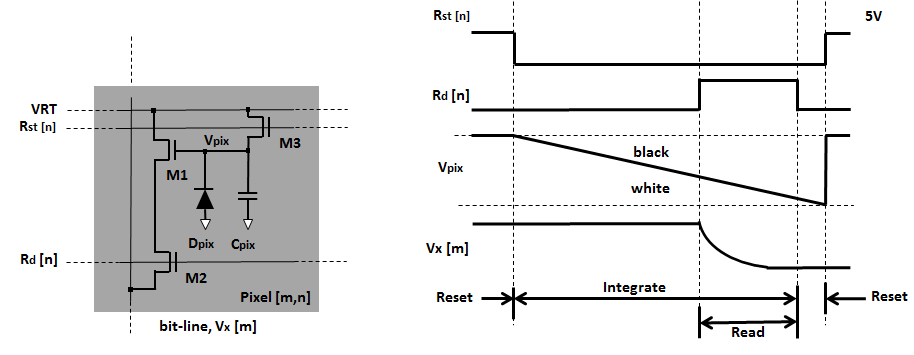}
\caption{Example of an active pixel along with graphs of voltage vs. time taken from various points within the pixel~\cite{Taylor}.}
\label{fig:CMOS1}
\end{figure}

In 1963 Frank Wanlass and Chih-Tang Sah from Fairchild Semiconductor invented logic circuits combining p-channel and n-channel MOS transistors in a complementary symmetry circuit configuration~\cite{Wanlass}. 
The patent describing the technology known as CMOS (complementary metal oxide semiconductor) was granted to Wanlass in 1967~\cite{Wanlass2} and the first CMOS image sensors were developed in late 60s. Although the CMOS sensors emerged almost simultaneously to CCD sensors, it took until 1993 when Jet Propulsion Laboratory produced the first CMOS sensor with performance comparable to the scientific-grade CCD sensor.
The underlying technology for CMOS sensors is quite different from the aforementioned CCD technology.

A CMOS pixel consists of a photo-diode, a capacitor, and transistors, as shown on the left side of Figure~\ref{fig:CMOS1}.
The CMOS imaging cycle can be roughly divided in three phases: integration, readout and reset. 
A typical CMOS imaging cycle is illustrated in Figure~\ref{fig:CMOS1}, right. 
At the beginning of an imaging cycle the capacitor is charged to a defined value. 
During the integration phase the capacitor is discharged via the photo-diode. 
The discharge is proportional to the incident photons.
In the next phase the remaining voltage is read out.
After the readout the capacitor is reset to the defined start value.

\begin{figure}[h!]
\centering
\includegraphics[width=0.8\textwidth]{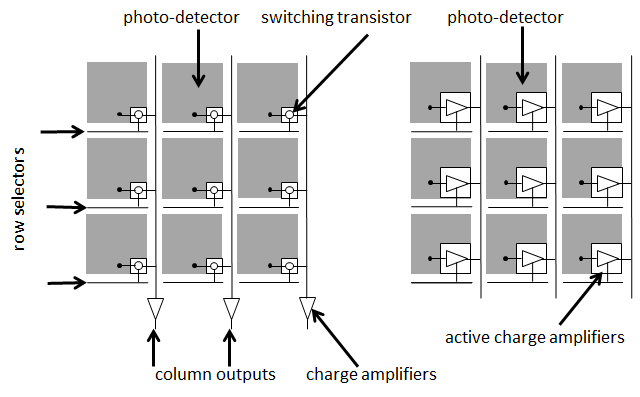}
\caption{Example of passive (left) and active (right) pixel CMOS array~\cite{Taylor}.}
\label{fig:CMOS2}
\end{figure}

CMOS sensors can be passive or active (Figure~\ref{fig:CMOS2}). 
In active picture sensors (APS) an amplifier is integrated in the pixel leading to better signal-to-noise ratio (SNR) and faster readout but smaller sensitivity due larger pixel size and smaller fill factor than passive pixels. 
The first active pixel sensor was created by Peter J.W. Noble in 1968~\cite{Noble}.

The CMOS technology enables to address each pixel for read out. 
Therefore various ways to read out CMOS imaging sensors are possible. 
Due to its smaller power consumption the rolling shutter readout is often preferred.

\subsection{CCD vs. CMOS}

\begin{figure}[h!]
\centering
\begin{subfigure}[b]{0.45\textwidth}
\centering
\includegraphics[width=0.8\textwidth]{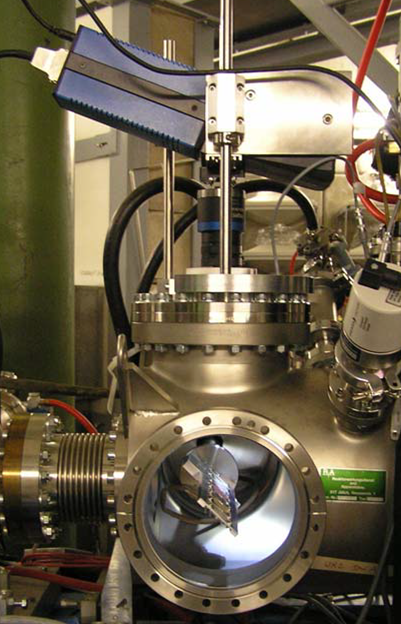}
\caption{}
\label{fig:CCDGSI}
\end{subfigure}
\begin{subfigure}[b]{0.5\textwidth}
\centering
\includegraphics[width=\textwidth]{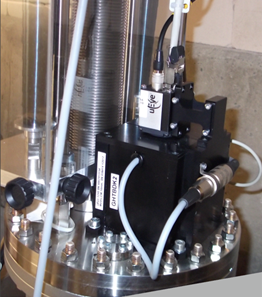}
\caption{}
\label{fig:CMOSGSI}
\end{subfigure}
\caption{Scintillating screen for beam profile measurements at the GSI linear accelerator using a CCD camera (a) and one equipped with a
CMOS camera in the high energy beam transfer line (b).}
\label{fig:GSICCDandCMOS}
\end{figure}
Both, CCD and CMOS sensors are commonly used in standard beam instrumentation applications (see Figure~\ref{fig:GSICCDandCMOS}).
CCD imaging sensors have technically developed over the last 50 years to the point where the high-resolution formats with very low defect rates are available. 
CMOS sensors have become dominant over CCD sensors in consumer products like video, smart phone, and digital camera applications due to their low cost and fast readout speed.

Although both technologies provide semiconductor based arrays of pixels, there are some differences.
CCD pixel act as capacitors and the signal is a charge which needs to be converted by an external circuit.
CMOS uses in-pixel photo-diodes and integrated circuits. 
The output is a voltage. 
The drawback of the integration of the circuits in the pixels of a CMOS sensor is an only moderate fill factor compared to CCD sensors.
Larger pixel size of CCD sensors results in better signal-to-noise ratio (see section~\ref{sec:SensorCharacteristics}).
CMOS sensors do not reach the high level of uniformity as CCD sensors, since the internal structure and therefore the fabrication process is more complex.
On the other hand, CMOS sensors are faster than CCD sensors and have a significantly lower power consumption.
Blooming as described in section~\ref{sec:SensorCharacteristics} does not effect CMOS pixel, since the signal is a voltage.
Integrated data acquisition and pre-processing circuits on the sensor make system complexity of CMOS based devices lower than for CCD based systems.
Both technologies are not radiation hard, although CMOS sensors are less susceptible to radiation. 
Table~\ref{tab:CCDvsCMOS} summarizes the comparison between CCD and CMOS sensors.

\begin{table}[h!]
\caption{CCD vs. CMOS.}
\label{tab:CCDvsCMOS}
\centering
\begin{tabular}{lllr}
\hline
    & CCD & CMOS \\
\hline
pixel signal & electrical charge & voltage \\
noise & low & moderate \\
fill factor & high & moderate \\
uniformity & high & slightly lower\\
speed & moderate & high \\
system complexity & high & low \\
power consumption & moderate & low \\
radiation & sensitive & less susceptible \\
\hline
\end{tabular}
\end{table}

Most CMOS sensors use rollings shutters which create unwanted artifacts if object and camera are moving fast relative to each other.
In the past only CCD sensors used global shutters delivering correct images of fast moving objects.
In the field of beam instrumentation, where pulsed light sources are common, one should take care of used shutter mechanism.
Today high quality CMOS sensors with global shutter are available.
A global shutter increases the number of transistors from two (rolling shutter) to four to five (global shutter). 
So that global shutter CMOS sensors have a reduced fill factor and an increased power consumption.

In March 2015, Sony announced that it was discontinuing its entire line of CCD sensors. 
This puts the end in sight for further development of CCD-based technology. 
Future sensors will be based on CMOS. 
CCD technology has still some advantages in large array sensors, but the gap continues to narrow.

\section{Image sensor: CID}
\label{sec:CID}
\begin{figure}[h!]
\centering
\includegraphics[width=0.55\textwidth]{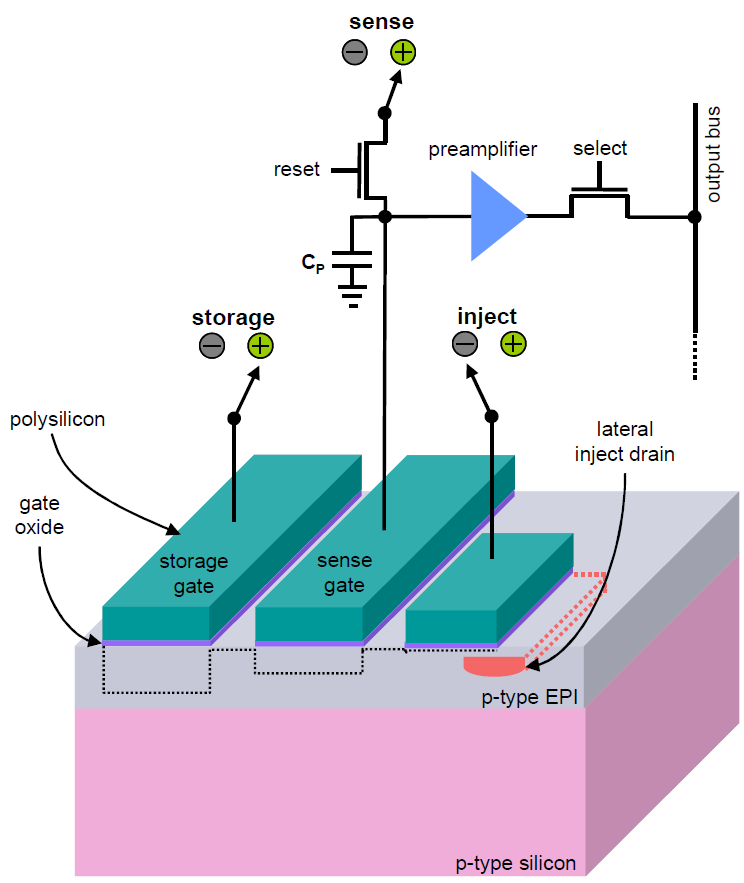}
\caption{Active pixel CID with storage and sense gatex~\cite{Bhaskaran}.}
\label{fig:CID2}
\end{figure}

\begin{figure}[h!]
\centering
\includegraphics[width=1\textwidth]{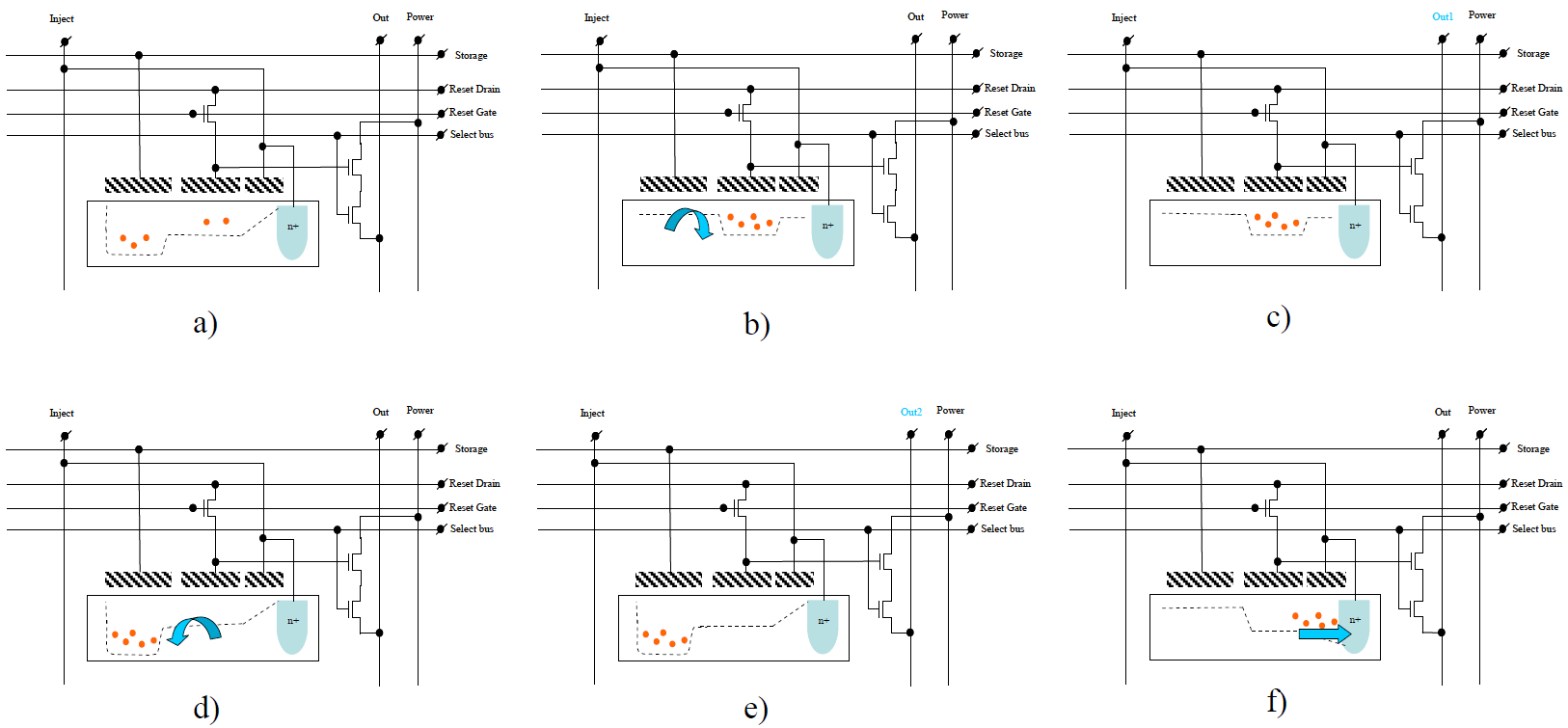}
\caption{The six phases of operation for active pixel CID. a) Integration, b) Forward-Transfer, c) Readout-Before-Transfer, d) Back-Transfer, e) Readout-After-Transfer  and f) Injection~\cite{Bhaskaran}.}
\label{fig:CID3}
\end{figure}

The Charge Injection Device (CID) was conceptualized by Hubert Burke and Gerald Michon at General Electric Company and evolved first into an  imaging sensor in 1972~\cite{Burke}. 
Like CCD, incident photons are converted to a proportional amount of charge, which is stored in the MOS capacitors of each pixel. In general, the CID pixel has two coupled MOS capacitors or two gates: ``storage`` and ``sense`` and a lateral inject drain as shown in Fig.~\ref{fig:CID2}. The photon-generated charge is accumulated under the ``storage`` gate.
The charge is measured by the voltage change induced by transferring the charge between the two gates. 
Figure~\ref{fig:CID3} shows the various phases of an active pixel operation and a detailed description can be found in~\cite{Bhaskaran}. After the integration/exposure stage, the charge is transferred to the sense gate. The sense gate is left floating using the reset FET, and sense gate voltage is measured. Then the charge is transferred back to storage gate and the sense gate is remeasured. The difference between these measurements is proportional to the amount of photon-generated charge within the pixel site. The photon generated charge may be cleared by collapsing the potential wells under both the ``storage`` and ``sense`` gates by ``injecting`` the charge into the underlying substrate to clear the array for new frame integration~\cite{Bhaskaran}.

Every pixel in a CID array can be individually addressed via electrical indexing of row and column electrodes. A displacement current proportional to the stored signal charge is read when charge "packets" are shifted between capacitors within individually selected pixels. The displacement current is amplified, converted to a voltage, and fed to the outside world as part of a composite video signal or digitized signal. Readout is non-destructive because the charge remains intact in the pixel after the signal level has been determined. 

Similar to CMOS, there  are  two  approaches  used  for  CID  image  sensors: passive pixel and active  pixel. Passive pixel is the simplest architecture with the largest fill factor, but has low conversion gain.
It is characterized by large full well up to 2 million electrons, high fill factor and very high radiation hardness. The active pixel, or preamplifier per pixel CID was developed for applications requiring higher dynamic range and established for improving imaging signal-to-noise ratio. Active pixel CIDs boast low readout noise, but the structure makes then complicated and sensitive to transient noise effects when exposed to gamma radiation~\cite{Bhaskaran}. 

\subsection{Comparison against CCD/CMOS}
Inspite of the increasing popularity of CMOS sensors for conventional imaging applications, CID-based cameras have found their niche in applications requiring extreme radiation tolerance and in high dynamic range scientific imaging. CIDs may be configured as a radiation hardened device that can reliably operate in a wide range of radiation environments well beyond the typical lifetime of CCD or CMOS-based cameras (in some cases orders of magnitude).
The operation is extended well into the MegaRad range by sensing the voltage threshold shifts that occur on the device due to radiation and dynamically adjusting the drive voltages to compensate for this increase in gate threshold~\cite{Chapman}. 

Other advantages of CIDs include sensitivity due to relatively high fill factors and a wide spectral response from 156\,nm to 1100\,nm. CIDs are also capable of true random access pixel addressing for nondestructive or destructive readout and charge clear. These capabilities allow for scientific images with extremely high dynamic range.

\section{Radiation Effects}
Materials used in a radiation environment can undergo degradation due to radiation damage effects. The negative effects of ionizing radiation on image sensors can be broadly classified into cumulative degrading total ionizing dose (TID) effect, displacement effects caused mainly by neutrons and statistical single events effects (SEE). The SI unit of absorbed dose is Gray equivalent to 1\,J/kg. 

\begin{figure}[h!]
\centering
\includegraphics[width=0.65\textwidth]{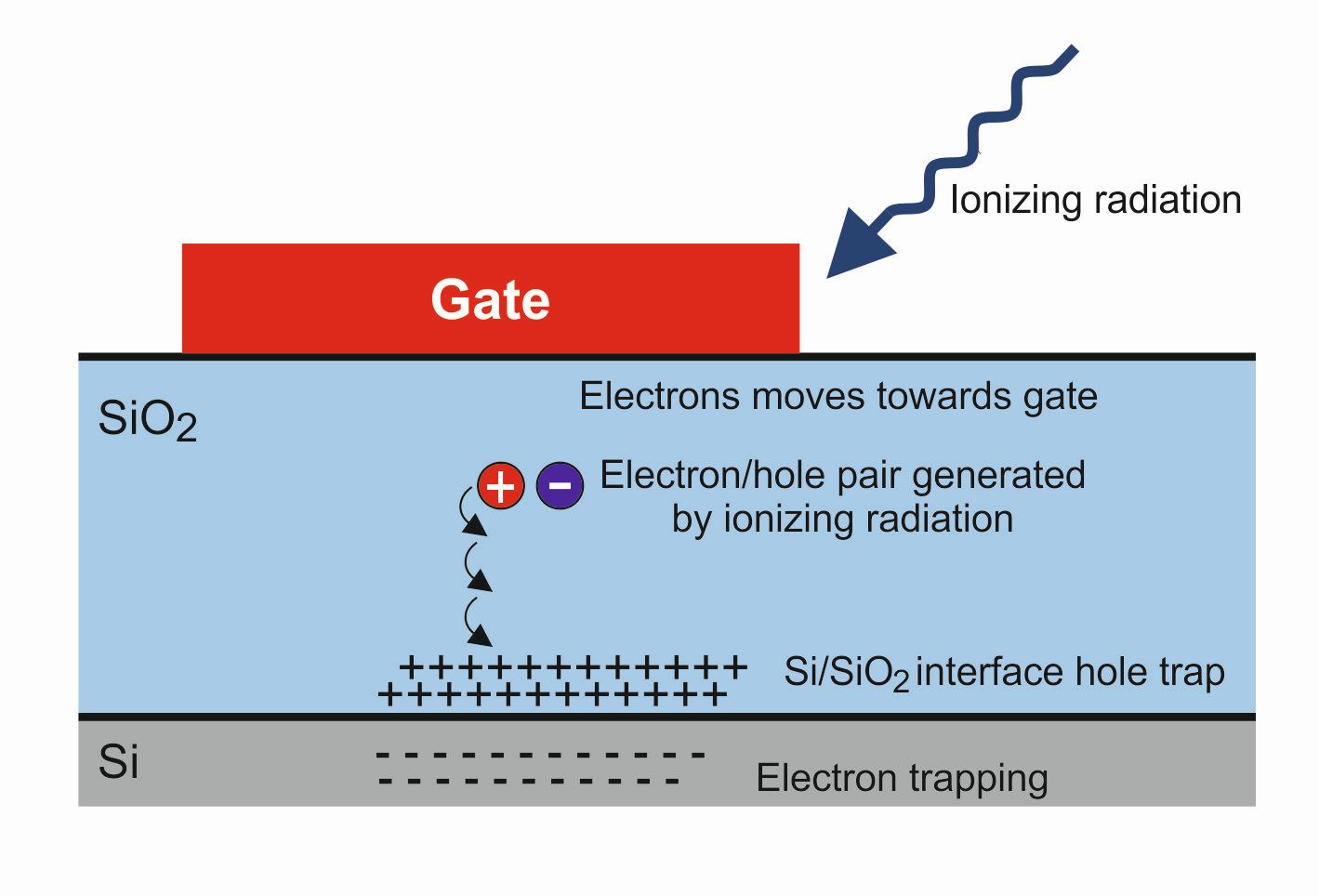}
\caption{Schema of the charge trapping on Si/SiO\textsubscript{2} interface layer~\cite{Kvedalen}.}
\label{fig:Rad1}
\end{figure}

Figure~\ref{fig:Rad1} shows the effects of radiation of  highly energetic particles or photons on a MOS capacitor. The most sensitive radiation effect is when the energy is deposited into the SiO\textsubscript{2} layer creating free electron-hole pairs. While some of the electrons and holes can recombine immediately i.e. in the first few nanoseconds, remaining free electrons will be attracted by the positive voltage on the gate. The less mobile holes move from their location towards the negative charged  Si-bulk after some time and will be trapped at the Si/SiO\textsubscript{2} interface. This positive charge attracts electrons in Si and cause current leakage leading to a higher power consumption. The positive charge trapped in SiO\textsubscript{2} layer will also create threshold voltage shifts. 

The holes trapped in the oxide layer are not stable and can disappear after some time due to effects referred to as annealing. Two types of this effect can be observed after the irradiation has been stopped: tunnel and thermal annealing~\cite{Kvedalen}. 
The annealing which takes place when the device is biased and operated at normal conditions is the tunnel annealing. The electrons close to Si/SiO\textsubscript{2} layer can recombine with holes in oxide layer due to quantum tunneling effect. The probability for this recombination strongly decreases with the distance from the Si/SiO\textsubscript{2} interface. A similar effect occurs on the gate-SiO\textsubscript{2} interface when the gate oxide is thin enough \cite{Kvedalen}. Thermal annealing occurs when the trapped holes are removed by increasing the temperature of the material up to 300$^o$C.

Displacement damages in semiconductors are usually caused by high energy neutrons which can dislodge or displace atoms in lattices and create vacancies. The neutron can excite an atom which then de-excites by emitting the photons and ionizes the semiconductor. Neutrons can also be absorbed by a target atom causing the formation of new atom in the silicon lattice. Protons or alpha particles created in this process can ionize the semiconductor and new atoms can increase resistivity of the material and reduce its lifetime~\cite{Kvedalen}.

The functionality of the system is not just dependent on the total dose absorbed, but also on the rate at which the dose was absorbed, the so called: dose-rate. The experiment~\cite{Burckhart} has shown that, for the same cumulative dose, low dose-rate irradiation caused an increase in the number of interface traps compared to circuits irradiated with a higher dose-rate.

During the irradiation, random single incidents called single event effects can occur. These kinds of errors are observed as "soft" if there is a bit flip in the memory cell, an I/O port or a change in some logic function or "hard" if it disrupts the functionality or worse, if the device malfunctions permanently. They are broadly categorized as follows~\cite{Kvedalen}:
\begin{itemize}
\item
Single Event Upset (SEU) is a general description of an incident in a circuit. The change can be in an analog circuit, digital circuit or in an optical component. 
\item
Single Event Latch-up (SEL) triggers a parasitic component in the circuit. This can lead to a rapid current increase, which can cause a malfunction of the device if the current is not limited. If the current increase does not damage the circuit, the SEL error can be removed by a power off-on cycle.
\item
Single Event Burnout (SEB) is an incident where the drain-source channel suffers from a burnout. This is a destructive error.
\end{itemize}

\begin{figure}[h!]
\centering
\includegraphics[width=0.8\textwidth]{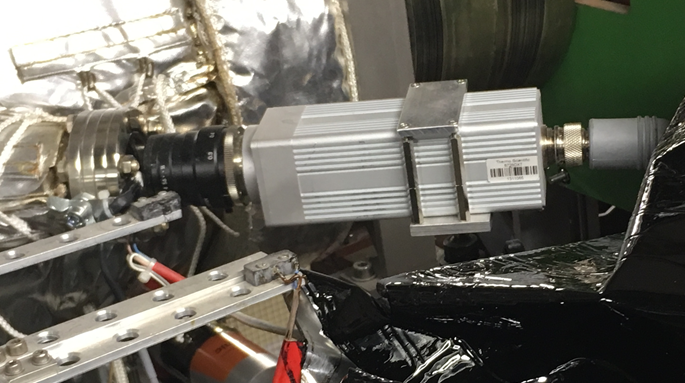}
\caption{Photo of MegaRAD3 CID camera with lens installed in GSI SIS18 extraction point.}
\label{fig:MegaRadGSI}
\end{figure}
\begin{figure}[h!]
\centering
\includegraphics[width=0.8\textwidth]{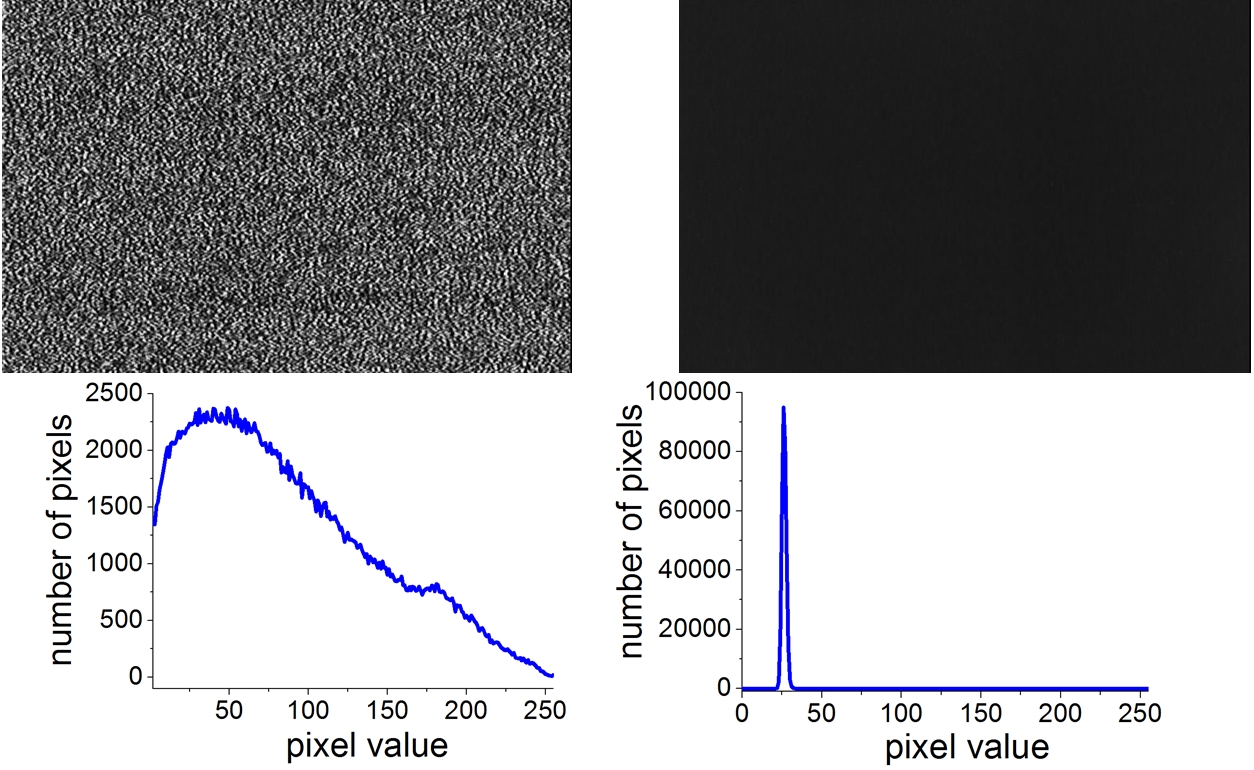}
\caption{Dark image of an standard CCTV camera (left) and radiation-hard CID camera (right) after few weeks of operation with corresponding histograms of brightness below \cite{Walasek2}.}
\label{fig:Rad2}
\end{figure}

The extraction point at the GSI synchrotron SIS18 is typically the location with high radiation level where 55\,kGy effective dose was recorded after eight months of beam operation. The MegaRAD3 camera was installed~\cite{Walasek2} at this site (see Figure~\ref{fig:MegaRadGSI}) and is continuously running without any failures for the past few years. In comparison, a standard CCD camera at the same position went out of order within two weeks. The dark image of standard CCD camera at a position with lower radiation level is shown in Figure ~\ref{fig:Rad2}, left. The CID-based camera exhibits a significant improvement for operation in a radiation rich environment as compared to the CCD and CMOS-based cameras. No significant change in the camera performance like quality of image, loss of contrast and resolution was observed.  
 This is inline with the tests performed by the manufacturer. These devices were found to be tolerant to gammas, neutrons, high energy electrons and proton radiation to at least 30\,kGy. First noticeable degradation in the image quality was reported for 140\,kGy exposure~\cite{Bhaskaran2}.

CCD and CMOS are typically not radiation hard; they can survive irradiation up to 100\,Gy. In this case CMOS is better due to the thinner oxide layers. For the video tube camera like Vidicon, there is no radiation limit as proven by many years of operation at CERN SPS~\cite{Bravin2} and only aging of the optics caused by irradiation can limit the system performance. Although, the tube-based cameras have to be protected from strong electrical and magnetic fields.

\section{Special Cameras}
\begin{figure}[h!]
\centering
\includegraphics[width=0.6\textwidth]{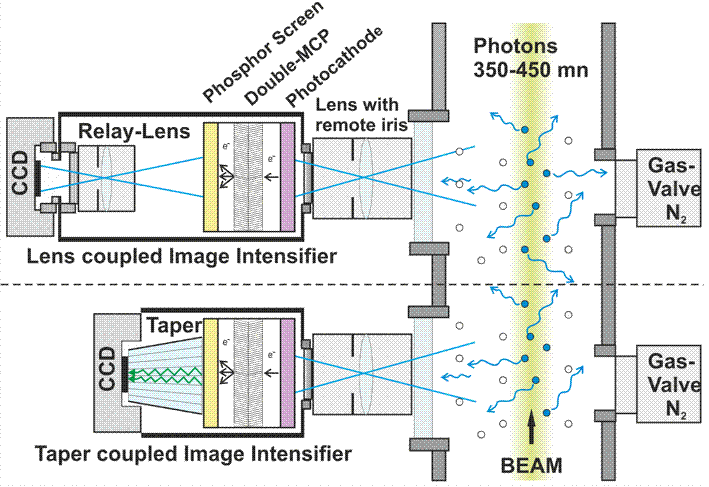}
\caption{Image Intensifier Camera systems used at GSI \cite{Andre}.}
\label{fig:BIF}
\end{figure}

In many special instrumentation applications, requirements on cameras become more challenging. When the intensity of available light is very low,  the back illuminated sensors~\cite{ALC} or electron multiplying CCD called emCCD~\cite{Hamamatsu2} cameras can be used. The emCCD achieved the high photo-sensitivity by an integrated electron multiplier. The electron multiplication takes place in several gain stages in shift register. Each of the stage multiples by an avalanche effect and the number of the stages determinate the gain factor. By integrating the amplifier stages into the charge coupled device, the inherent noise is low and the signal-to-noise ratio improves with the multiplication.
In the Ionization Profile Monitors~\cite{Giacomini} and Beam Induce Fluorescence~\cite{Andre} the single photon detection is must. In this case, an image intensifier device can be used to increase the signal strength. This device converts light photons into electrons with a photocathode which releases electrons via photoelectric effects. The electrons are accelerated through high voltage potential into a micro channel plate (MCP). Each electron releases multiple electrons. The MCP is made out of many tiny conductive channels which are tilted to encourage more electron collisions and thus enhance the emission of secondary electrons. The two-stage MCP can have multiplicity factor of $10^6$. Due to the high voltage the secondary electrons  move out of MCP and hit a phosphor screen at the other end of the intensifier, which release a photon for every electron. The image on the phosphor is coupled to the standard camera system with a fiber-taper or relay lens as illustrated at Figure~\ref{fig:BIF}.

\begin{figure}[h!]
\centering
\includegraphics[width=1\textwidth]{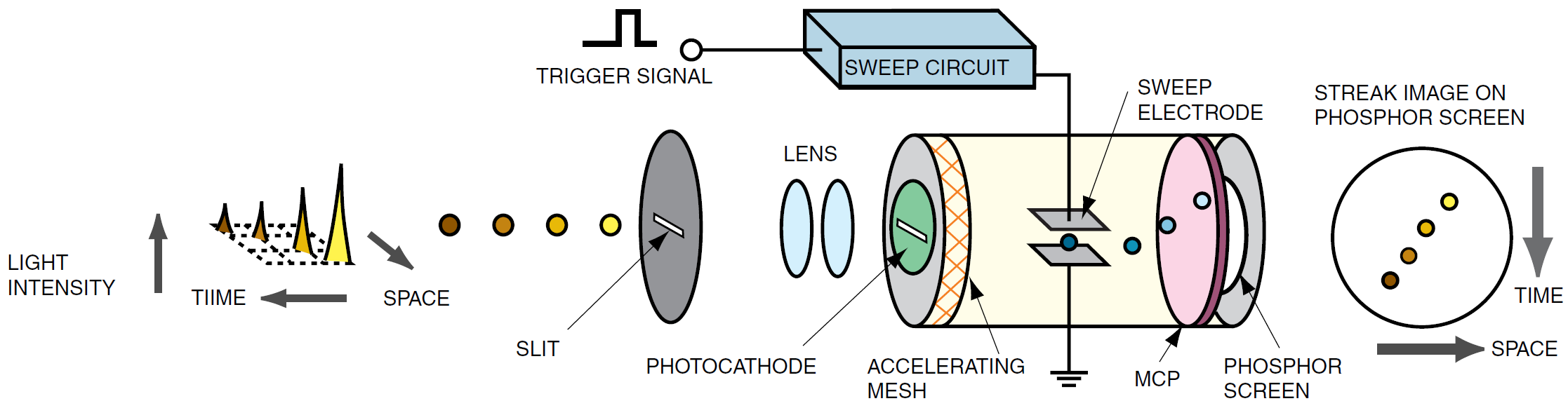}
\caption{Operating principle of the streak tube ~\cite{Streak}.}
\label{fig:Streak}
\end{figure}

In some instrumentation applications where frame rates of kHz or better are required, high speed cameras~\cite{Shimamoto} permit precise measurements and instructive visualizations of beam behaviour that cannot be obtained from other instrumentation devices. In many synchrotron light sources, ultra high speed detectors which capture light emission phenomena occurring in an extremely short time are required for the measurements of bunch length or ultra fast transverse motion~\cite{Scheidt}. The streak camera (Figure~\ref{fig:Streak}) transforms the time variations of the very short light pulse into a spatial profile. A light pulse enters the streak unit through a narrow slit, hits the photocathode, and releases electrons. These electrons are accelerated and get deflected by a sweep electrode in the perpendicular direction (from top to bottom) so that the electrons that arrive first hit the MCP at a different position in comparison to electrons that arrive later. A phosphor screen converts electrons back into light which is then coupled to a standard camera. 

For the electron beams, the usage of the beam imaging techniques was extending to perform precise  measurements of important beam parameters such as emittance, energy, energy spread by using for the optical transition radiation (OTR) as reported in \cite{Ischebeck}. 

\section{Data Acquisition, Information Transfer and Post Processing}

\subsection{Image Digitization}
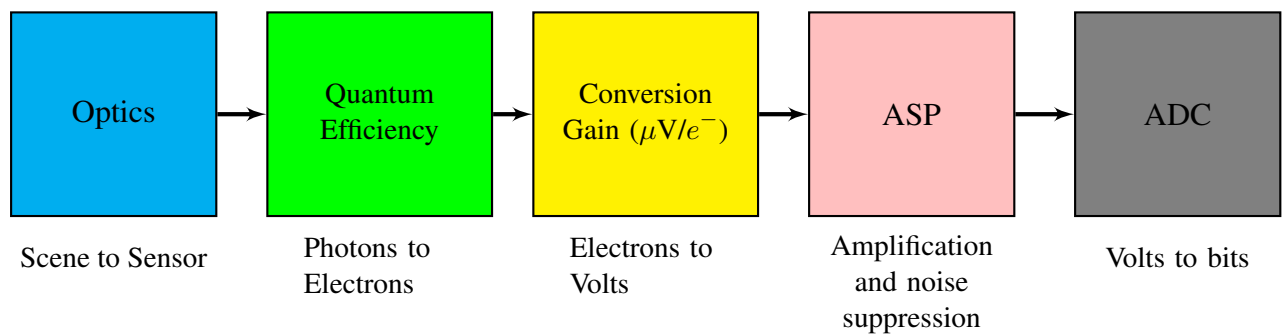
\begin{figure} [h!]
\centering
\begin{tikzpicture}
\node[thick,draw,minimum width=2.7cm,minimum height=2.7cm,fill=cyan] (Optics) at (0,3) {\large{Optics}};
\node[thick,minimum width=2.7cm] (label1) at (0,1.1) {Scene to Sensor};
\node[thick,draw,text width=2.7cm,align=center,minimum height=2.7cm,fill=green] (QE) at (3.5,3) {Quantum Efficiency};
\node[thick,text width=2cm] (label2) at (3.5,1) {Photons to Electrons};
\node[thick,draw,text width=2.7cm,align=center,minimum height=2.7cm,fill=yellow] (CG) at (7,3) {Conversion Gain ($\mu$V/$e^-$)};
\node[thick,text width=2cm] (label3) at (7,1) {Electrons to Volts};
\node[thick,draw,minimum width=2.7cm,minimum height=2.7cm,fill=pink] (ASP) at (10.5,3) {\large{ASP}};
\node[thick,text width=2.7cm,align=center] (label1) at (10.5,0.75) {Amplification and noise suppression};
\node[thick,draw,minimum width=2.7cm,minimum height=2.7cm,fill=gray] (ADC) at (14,3) {\large{ADC}};
\node[thick,text width=2.7cm] (label1) at (14.4,1.1) {Volts to bits};
\draw[ultra thick,-latex'] (Optics)--(QE);
\draw[ultra thick,-latex'](QE)--(CG);
\draw[ultra thick,-latex'](CG)--(ASP);
\draw[ultra thick,-latex'](ASP)--(ADC);
\end{tikzpicture}
\caption{Steps from photons to bits \label{fig:Dig1}}
\end{figure}

In order to store and post-process information transmitted by light it has to be digitized.
The camera optics guides the photons on the image sensor.
Photons in the range of visible light create electron-hole pairs in silicon based semiconductors.
Image sensor chips use this process to convert the information transmitted by the photons to charge (see Sections~\ref{sec:CCD}, \ref{sec:CMOS}, \ref{sec:CID}). 
The quantum efficiency (see Section~\ref{sec:SensorCharacteristics}) is a measure of the information loss during this process.
The charge is converted into a voltage to read out the signal.  
The conversion gain determines the amount of Volts per electron for the pixels. 
Typical values range from 1 to 10\,$\mu$V/e. 
In CMOS image sensors the conversion chain photon - electron-hole pairs - voltage is integrated in the pixel. 
For CCD and CID sensors the charge to voltage conversion is outside of the pixels.
The analog voltage signal is digitized with an analog-to-digital converter (ADC). 
The ADC is not part of the image sensor. 
ADC can have an analogue signal processor (ASP) to reduce noise and provide additional gain before conversion. 

\begin{figure}[h!]
\centering
\includegraphics[width=0.9\textwidth]{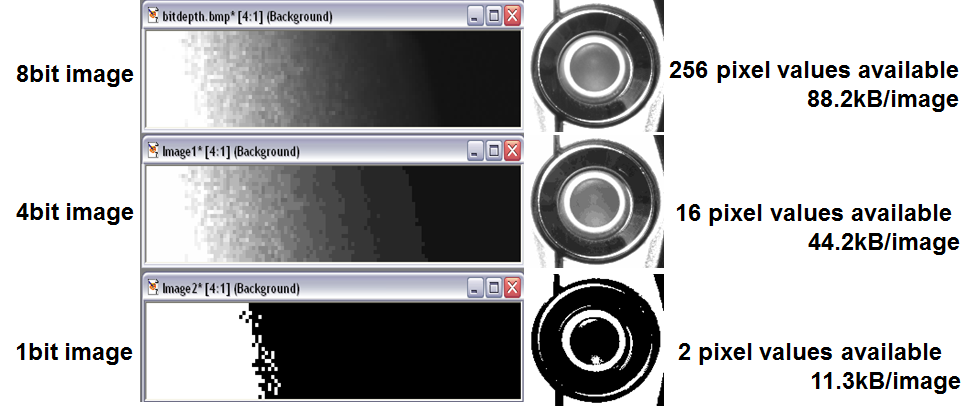}
\caption{Image taken with different dynamic range: 8\,bit (top), 4\,bit (middle) and 1\,bit (bottom) \cite{Chouinard}.}
\label{fig:Dig4}
\end{figure}

The ADC quantizes the interval from minimum to maximum input voltage into discrete values.
The possible values are powers of 2. 
In the ADC community powers of 2, expressed in bits, is known as the resolution of the ADC. 
In the imaging sensor community this value is named bit depth or dynamic range.
A dynamic range of 1\,bit results always in a black and white image (see Figure~\ref{fig:Dig4} bottom), whereas a dynamic range of 8\,bit provides images with up to 256 shades of gray (see Figure~\ref{fig:Dig4} top).
The higher the dynamic range the more information is in the captured image, but its file size increases, too. 

The untrained human eye distinguishes approximately 256 shades of gray. 
Standard displays and TV screens are adapted to this.
Trained eyes may see 10\,bit information. 
But still sometimes a higher dynamic range of 12\,bit or even 16\,bit might be useful, since post-processing can reveal former hidden information. 
On the other hand, low dynamic range images require less bandwidth and those sensors are cheaper. 

\subsection{Cameras Interface Standards}

Early video cameras used an analogue interface. This interface enables easy real-time viewing but makes it difficult to capture and store images for subsequent digital post-processing. To convert TV signals into digital information, a special device was developed called a frame grabber. This board ``grabs`` images for digital storage.
Similar to the well known analog video signal standards RS 170/CCIR, NTSC, and PAL digital video interface standards were developed starting in the mid 1990s. 

In 1995, Institute of Electrical and Electronics Engineers (IEEE) introduced IEEE 1394 FireWire as a new interface. The FireWire interface made transferring image data from a camera to a computer easy and cost-effective.
In October 2000, Automated Imaging Association (AIA) introduced a digital interface called Camera Link. Camera Link simple hardware reduces camera costs but increases system costs, as it requires the use of special frame grabber boards and demanding high quality transmission cables to operate.
USB technology has been available in Version 2.0 since 2000 and provides sufficient bandwidth for small resolution cameras. USB took a big step forward with USB 3.0 in 2008. This USB interface allows the delivery of power via the cable. 
The development of Gigabit Ethernet (GigE) in 1999 makes it possible to connect several cameras on a single port using network switches and provides power over the cable (Power over Ethernet PoE). In 2006, GigE Vision became the AIA standard control and image transmission protocol.
The foundation for a generic camera interface called GenICam in 2006, administrated by the European Machine Vision Association (EMVA), was a big step forward to improve compatibility over camera manufacturers proprietary protocols into common control and data transmission protocol.  
In 2012, a new standard called CoaXPress was introduced. The advantage of the CoaXPress is the possibility to use a single coax cable for camera control, image data transmission, power supply, and camera trigger.

The choice of the interface standard depends on the demand in each individual case. 
Quality and frequency of the grabbed frames define the needed bandwidth. 
In an accelerator facility operating and radiation safety aspects or simple space requirements define minimum cable length and thus influence the choice of the interface standard.
Last but not least, the choice should be made according to the standards used in the facility.
Table~\ref{tab:interfaces} summarizes the interface standards, their bandwidth and the maximum cable length.
More detailed information is given e.g. in \cite{PCO}.

\begin{table}[h!]
\caption{Data interfaces for image data transfer with bandwidth and cable length.}
\label{tab:interfaces}
\centering
\begin{tabular}{lllr}
\hline
 Standard & Bandwidth & Cable length \\
\hline
FireWire - IEEE 1394a & 400\,Mbits/s & 10\,m \\
FireWire - IEEE 1394b & 800\,Mbits/s & 10\,m \\
Gigabit Ethernet - GigE & 1\,Gbits/s & 100\,m \\
10~Gigabit Ethernet~-~10GigE & 10\,Gbits/s & 100\,m \\
USB 2.0 - Speed USB & 480\,Mbits/s & 3 - 5 m \\
USB 3.0~-~SuperSpeed USB & 3.5\,Gbits/s & 3~-~5\,m \\
CoaX Press & 6.25\,Gbits/s & 100\,m \\
Camera Link & 6.8\,Gbits/s & 10\,m \\
Camera Link HS & 16.8\,Gbits/s & 10.000\,m \\
\hline
\end{tabular}
\end{table}

\subsection{Post processing}
In beam instrumentation images need to be analysed in order to extract the useful beam parameters like position and width of the beam spot. The first steps usually consist of choosing the correct region-of-interest and eliminating the unwanted features from the image like: background noise, reflections and fixed patterns. Online processing can include cropping, rotation (90$^\circ$, 180$^\circ$ and 270$^\circ$), flipping and simple image scaling (conversion of pixels into millimetres). Typical online analysis consists of creating projections in x and y direction as well as the intensity histogram. Center-of-brightness and FWHM can be calculated based on the projections. 

\begin{figure}[h!]
\centering
\includegraphics[width=1\textwidth]{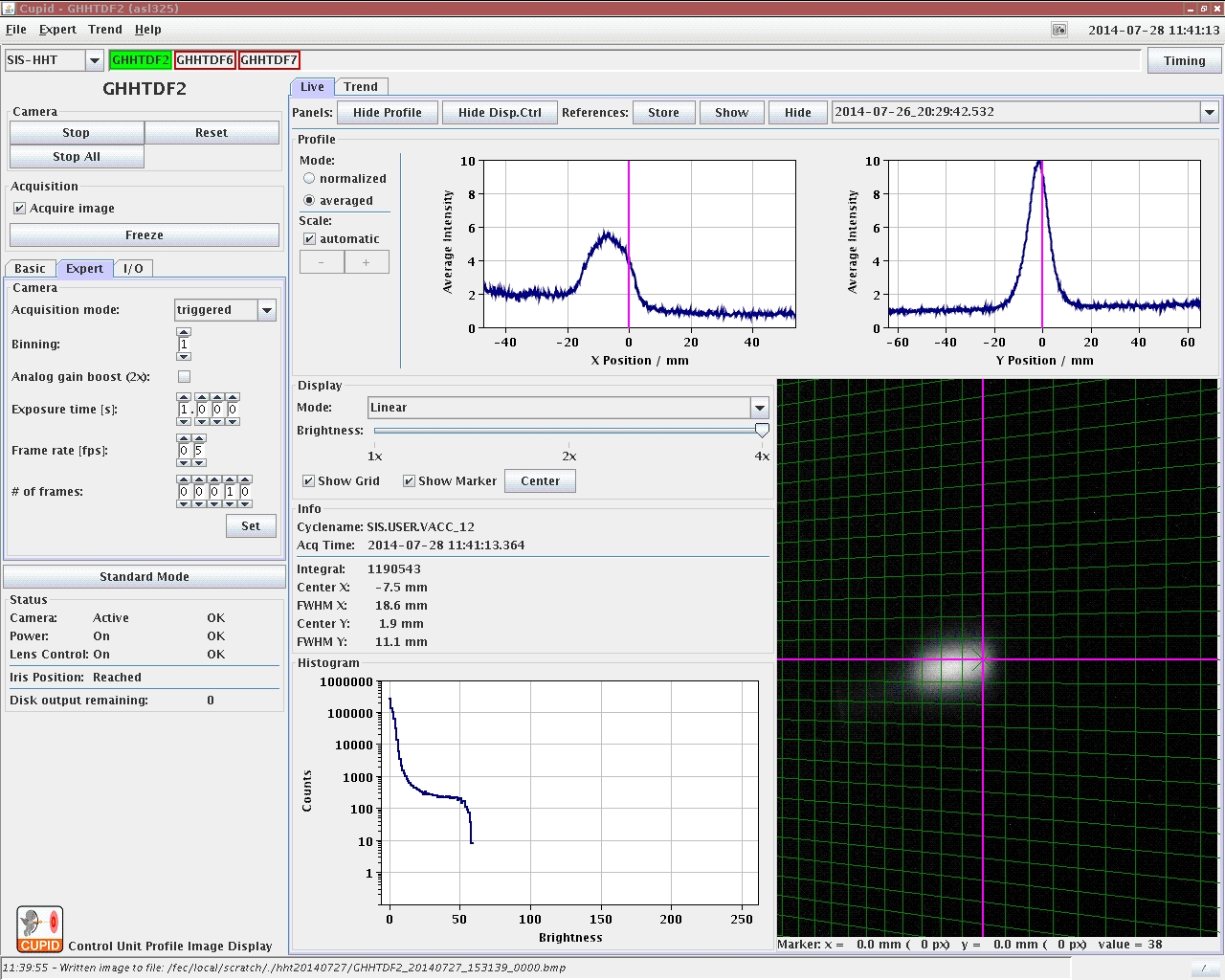}
\caption{An example of graphical user interface (GUI) as provided at GSI. The camera control panel, image display and its analysis are shown.}
\label{fig:GUI}
\end{figure}

For the basic beam instrumentation operation the camera controls are reduced to changing the exposure time, the binning of the image or requesting the raw images to be automatically saved to disk. It also allows changing the acquisition mode. The two most important ones are 'free run' and 'triggered'. In 'free run' the camera continuously acquires images with the specified exposure time and frame rate. The acquired images are displayed as they arrive in real time. In the 'triggered' mode, the image acquisition is triggered by a machine event of the accelerator (for example at beam extraction). At the time of the trigger, a single image is acquired and displayed. An extension of the 'triggered' mode is the 'sequence' mode, which acquires a pre-defined number of images with the specified frame rate after a trigger is received.
Recently at GSI, the readout of analog cameras like the CCTV type CCD, Thermo Fischer CID or Vidicon has been implemented using the Pleora iPort Analog-Pro external frame grabber. This module is Gigabit ethernet based and compatible with the GigE Vision and GenICam standards. As analog cameras have a fixed frame rate, a rate reduction in software to 10\,Hz reduces the amount of transported data. In addition a software trigger is implemented by discarding all frames but the one directly following a machine event detected by the data acquisition software.

\section{Conclusions}

In the last decades, imaging sensors have found an increasing number of applications in accelerator facilities. For standard scintillating screen applications, nearly all the camera types are used, with consideration of the radiation levels in the operation area. The tube camera like Vidicon are truly radiation hard imaging system, with lower resolution, sensitivity and signal to noise ratio compared to solid state systems. 

The common basis of CCD and CMOS technologies is the photosensitive elements that collects the charges liberated by the photons. The way the charge is read out differs then from one technology to the other.
In CCD sensors the charge is first simultaneously shifted into a storage area, the charges of each column of the storage area are shifted one by one toward the read out register. On its turn the cells of the read out register is shifted horizontally to the ADC. In this way all the pixels are digitized by the same ADC and it is relatively easy to create a uniform response over the whole sensor area. The disadvantage is that the readout is slow and the production is costly as it requires high grade silicon. The use of high grade silicon result in low dark current so long exposures are possible.
The CCD was the leading technology for scientific use for a long time, as in beam instrumentation. The main CCD producer, Sony, however has announced the discontinuation of the CCD production in favour of CMOS sensors in 2015.

CMOS sensors have been around for as long as CCDs but until recently were low cost, low performance sensors. The CMOS technology has been improved greatly and CMOS sensors have now matched and in some cases surpassed the performance of the CCD sensors. The CMOS sensors are based on a complex pixel design with usually up to 5 transistors (3 transistors for the base CMOS design). The complex design of the CMOS pixel reduces the  fill factor of the pixel. This effect can be reduced by using micro lenses on each pixel to focus the light on the sensitive part of the pixel or by illuminating the sensor on the back side. Present state-of-the-art CMOS sensor offer good performance also in beam instrumentation.

CID technology offers advantages over both CMOS or CCD-based camera systems wherever radiation hardness  and high dynamic range is required. Both these issues form the core of beam instrumentation requirements. 
 
On the image quality side, the commercial market of solid state sensors is evolving extremely fast with a trend of providing more signal processing immediately after imaging. 
Here, the issue of radiation damage is still conflicting with placement of more ``intelligence`` in the camera sensor and head.

Experiences between accelerator labs are still being shared on this topic of "Imaging for beam instrumentation".
Recently a Joint ARIES-ADA Workshop on ``Scintillating Screens and Optical Technology for transverse Profile Measurements``,  was held \cite{Workshop} where many relevant challenges were discussed and reader will find current challenges and trends on this topic.

\section{Acknowledgement}
This proceeding condenses the experience accumulated over the years on the optical monitors in industry and accelerator facilities. The authors would like to thank the many colleagues in laboratories world wide for sharing their experience. It is a pleasure to acknowledge our colleagues Harald Braeuning and Hansi Roedl for being available for discussion and Aditya Tripathi for proofreading this contribution.


\end{document}